\documentclass[3p,times]{elsarticle}

\usepackage{ecrc}
\usepackage{subfigure}
\usepackage{wrapfig}

\volume{00}

\firstpage{1}

\journalname{Physics Procedia}

\runauth{}

\jid{phpro}

\jnltitlelogo{Physics Procedia}

\usepackage{amssymb}
\usepackage[figuresright]{rotating}

\begin{document}

\begin{frontmatter}



\dochead{}

\title{Concept and status of the CALICE analog hadron calorimeter engineering prototype}


\ead{mark.terwort@desy.de}
\author{Mark Terwort on behalf of the CALICE collaboration}

\address{DESY, Notkestrasse 85, 22607 Hamburg}

\begin{abstract}

  A basic prototype for an analog hadron calorimeter for a future
  linear collider detector is currently being realized by the CALICE
  collaboration. The aim is to show the feasibility to build a
  realistic detector with fully integrated readout electronics. An
  important aspect of the design is the improvement of the jet energy
  resolution by measuring details of the shower development with a
  highly granular device and combining them with the information from
  the tracking detectors. Therefore, the signals are sampled by small
  scintillating tiles that are read out by silicon photomultipliers.
  The ASICs are integrated into the calorimeter layers and are
  developed for minimal power dissipation. An embedded LED system per
  channel is used for calibration. The prototype has been tested
  extensively and the concept as well as results from the DESY test
  setups are reported here.

\end{abstract}

%%\begin{keyword}




%%\end{keyword}

\end{frontmatter}

% \linenumbers

\section{Introduction}
\label{}

The CALICE collaboration~\cite{CALICE} is currently developing a new
engineering prototype~\cite{Reinecke} of the analog hadron calorimeter
(AHCAL) option for a future linear collider (LC) experiment. A major
aspect of the design is the improvement of the jet energy resolution
by measuring details of the spatial development of all hadronic
showers inside a jet. These information are combined with the
information obtained from the tracking systems in order to measure
only the energies of neutral particles with the calorimeter system.
This concept is known as {\it particle flow} and has been
validated~\cite{PF} with the physics prototype~\cite{PPT} of the
CALICE AHCAL. The aim of developing an engineering prototype is to
demonstrate that a scalable device can be built that meets the
reqirements of a realistic LC detector, such as fully integrated
front-end electronics in the active layers of the calorimeter. It is
based on scintillating tiles, that are read out by silicon
photomultipliers (SiPMs).

First subunits (HCAL base unit, HBU) with 144 detector channels of
size $36\times 36$\,cm$^2$ have been produced and extensively tested
in the laboratory as well as in the DESY test beam
facility~\cite{Reinecke, Jeremy, Ivan}. They include the scintillating
tiles, four front-end low power dissipation SPIROC2 ASICs~\cite{ASICs}
and the light calibration and gain-monitoring system. The first
version of the detector/DAQ interface modules is used for power supply
and slow control programming. For studying a realistic LC operation
mode the power supply module allows for switching off individual
detector components within the LC bunch train structure ({\it power
  pulsing})~\cite{Peter}.

Recently all sub-components of the HBU (tiles, ASICs, calibration
system) as well as the DAQ modules have been redesigned in order to
optimize the performance and the spatial dimensions. First tests have
been performed with the next generation SPIROC2b ASIC~\cite{Oskar}. In
this report the concept and the status of the design as well as
results from test measurements are presented.

\section{Design and status of the engineering prototype}

\begin{figure}
\centerline{\includegraphics[width=0.7\columnwidth]{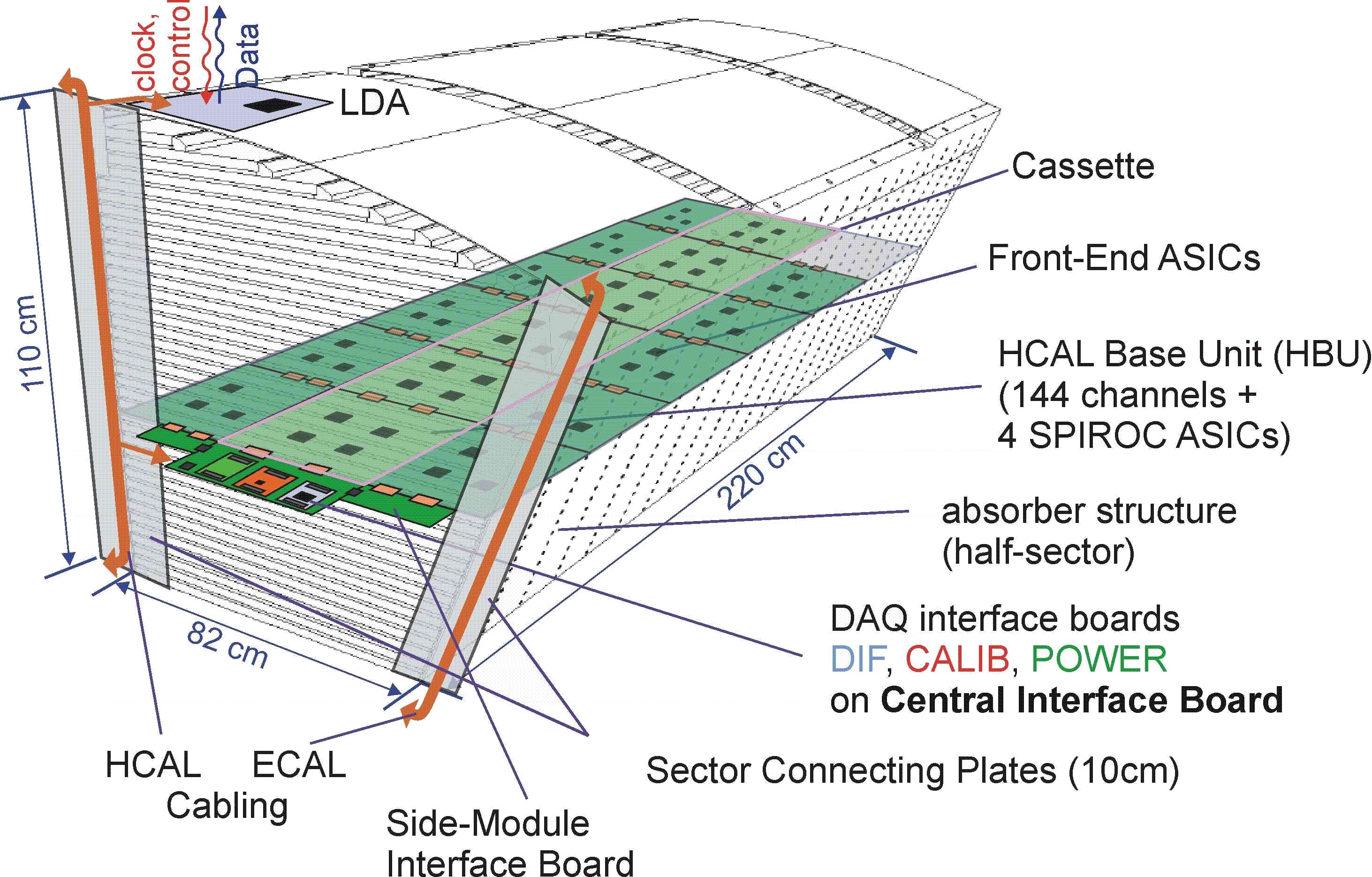}}
\caption{1/16 HCAL half-barrel with 48 layers containing three
  slabs each. The DAQ interface modules are also shown.}
\label{halfsector}
\end{figure}

The design of one half-octant of the AHCAL is shown in
Fig.~\ref{halfsector}. It consists of 48 layers, giving a total
thickness of 110\,cm. The barrel is divided into two sections, which
have a length of 220\,cm each. One layer consists of the 16\,mm thick
stainless steel or 10\,mm thick tungsten absorber plate and an active
layer part. This active part consists of the scintillating tiles with
an attached SiPM as photo detector and the embedded front-end
electronics with the readout chips and the calibration system. Each
layer has about 2500 channels, which adds up to about 4 million
channels for the whole barrel calorimeter.

\subsection{HCAL base units}

\begin{wrapfigure}{r}{0.4\columnwidth}
\centerline{\includegraphics[width=0.35\columnwidth]{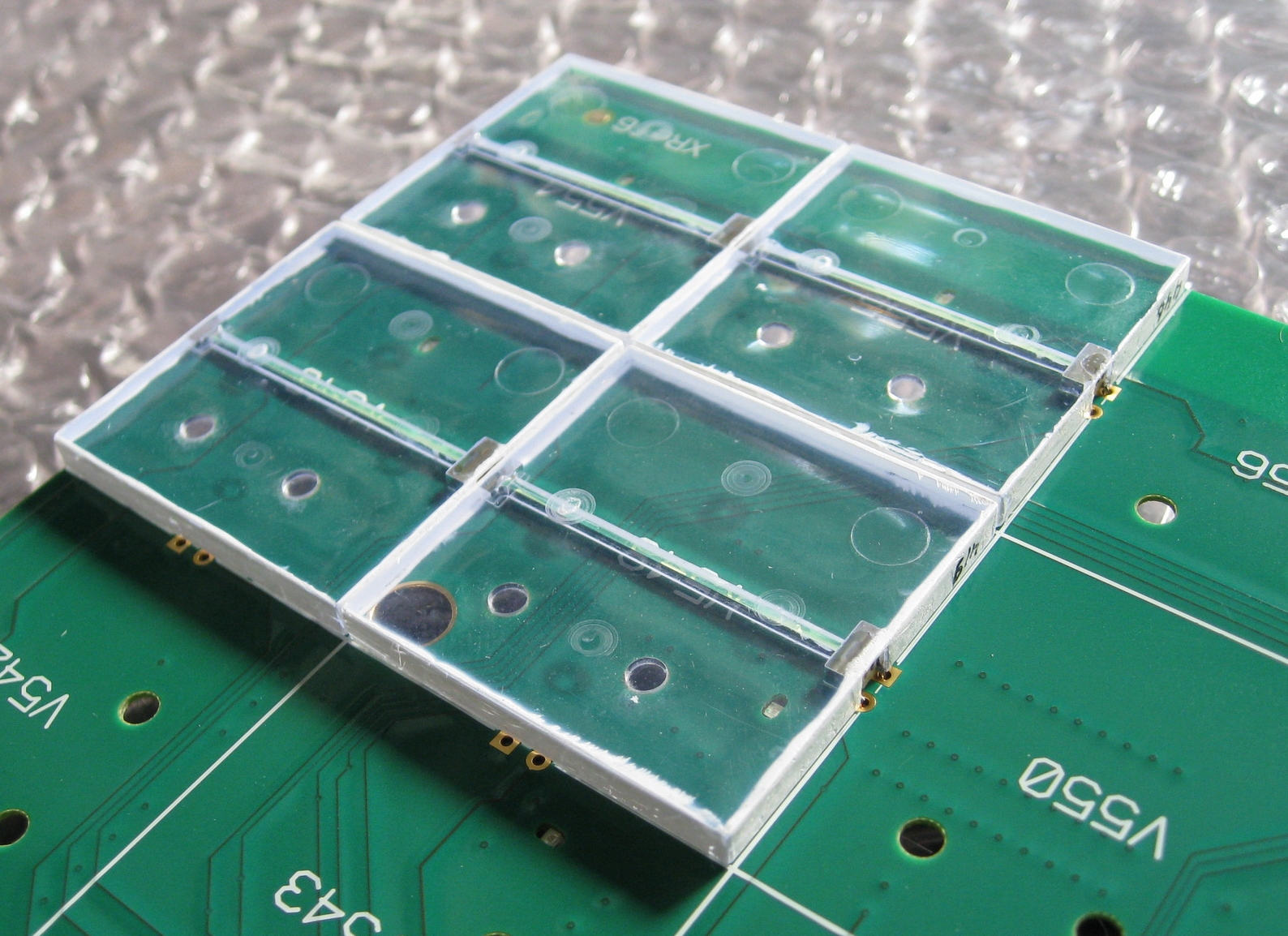}}
\caption{Scintillating tiles with embedded wavelength shifting fiber,
  SiPM, mirror and alignment pins as they are assembled below the
  PCB.}
\label{Tile}
\end{wrapfigure}

A single layer is divided into three parallel slabs, each of which
consists of six HBUs that are interconnected by ultra-thin flexleads.
The first version of the HBU has been tested extensively at DESY with
charge injection and LED light as well as in the DESY electron test
beam environment~\cite{Reinecke}. It features 144 detector channels of
$3\times 3$\,cm$^2$ size. The signal for each channel is produced by a
scintillating tile with 3\,mm thickness, that includes a straight
wavelength shifting fiber coupled to a SiPM with a size of
1.27\,mm$^2$ on one side and to a mirror on the other side. The SiPM
comprises 796 pixels with a gain of $\sim 5\cdot 10^6$. The tiles are
connected to the PCB by two alignment pins that are plugged into holes
in the PCB, while the nominal tile distance is 100\,$\mu$m. A photo of
the actual tile assebly is shown in Fig.~\ref{Tile}.

\begin{wrapfigure}{r}{0.4\columnwidth}
\centerline{\includegraphics[width=0.35\columnwidth]{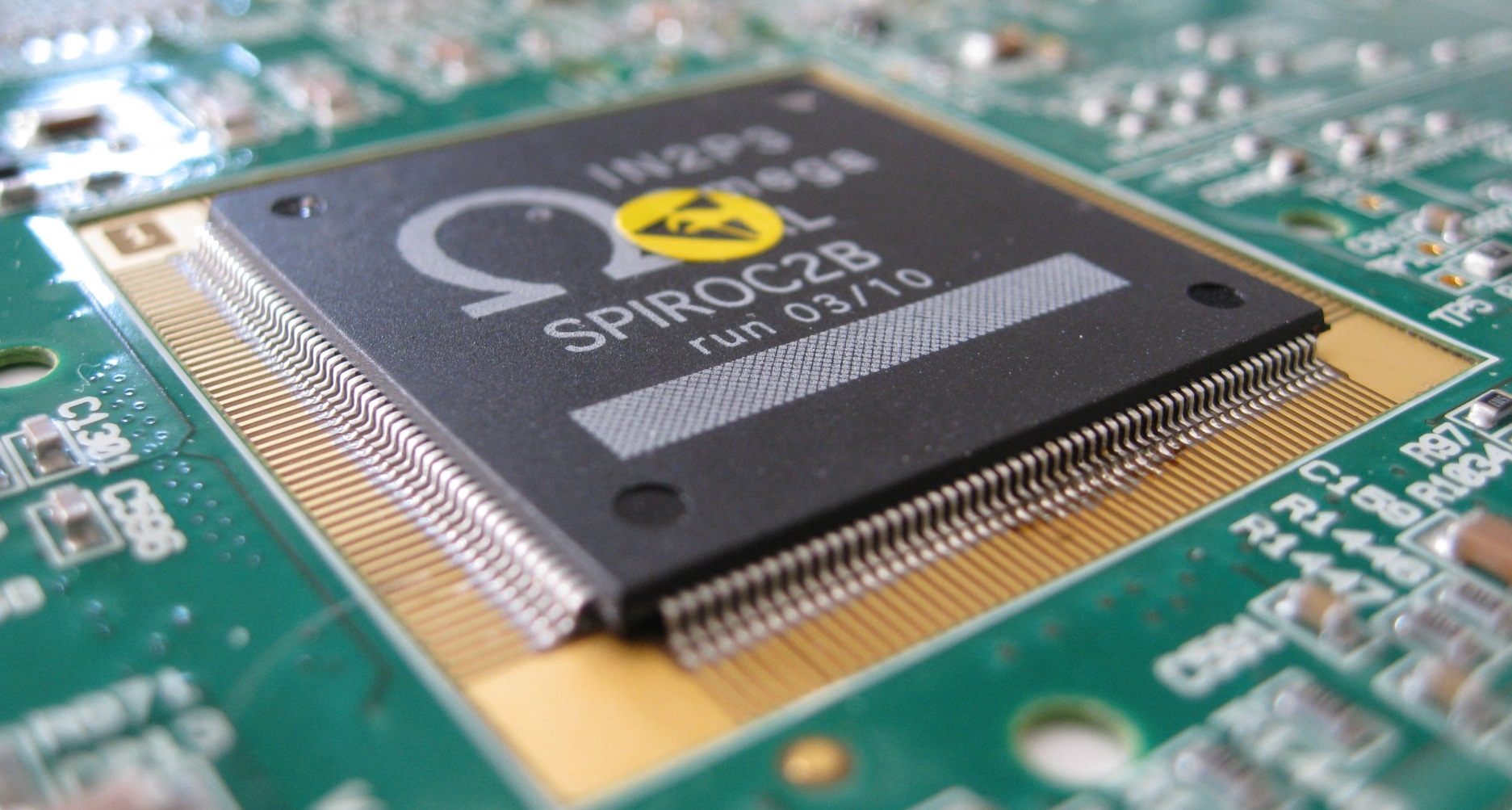}}
\caption{Integration of the SPIROC2b ASIC into the PCB.}
\label{SPIROC2b}
\end{wrapfigure}

The analog signals of the SiPMs are read out by 36-channel
ASICs~\cite{ASICs} equipped with 5\,V DACs for channel-wise bias
voltage adjustment. They provide two gain modes, where the high gain
mode is maily used as a calibration mode and the low gain mode
measures signals with higher amplitudes up to SiPM saturation. The
foreseen power consumption amounts to 25\,$\mu$W (40\,$\mu$W) per
channel (including SiPM bias) for the final LC operation in order to
avoid the need for active cooling. This means that parts of the chip
have to be switched off according to the bunch train structure of the
LC to save power. First tests of the performance of the power pulsing
mode in a full HBU setup have been performed with the SPIROC2 chip,
which gave good results~\cite{Peter}. The ASICs include the
digitization step (12-bit ADC and 12-bit TDC for charge and time
measurement) and the on-detector zero suppression with an adjustable
threshold. This self-triggering capability has been extensively tested
in the DESY test beam environment~\cite{Jeremy}. Additionally, the
chip can automatically chose between high gain and low gain channel,
which is described in Sec.~\ref{AG}. To reduce the height of the
active layers the ASICs are lowered into cutouts of the PCB by
$\sim$500\,$\mu$m. Recently a new version of the ASIC, the SPIROC2b,
has become available, mounted on the first version of the HBU and
first tests have been performed. A photo of the new chip as it is
integrated into the HBU is shown in Fig.~\ref{SPIROC2b}.

\begin{figure}[b!]
\centerline{\includegraphics[width=0.6\columnwidth]{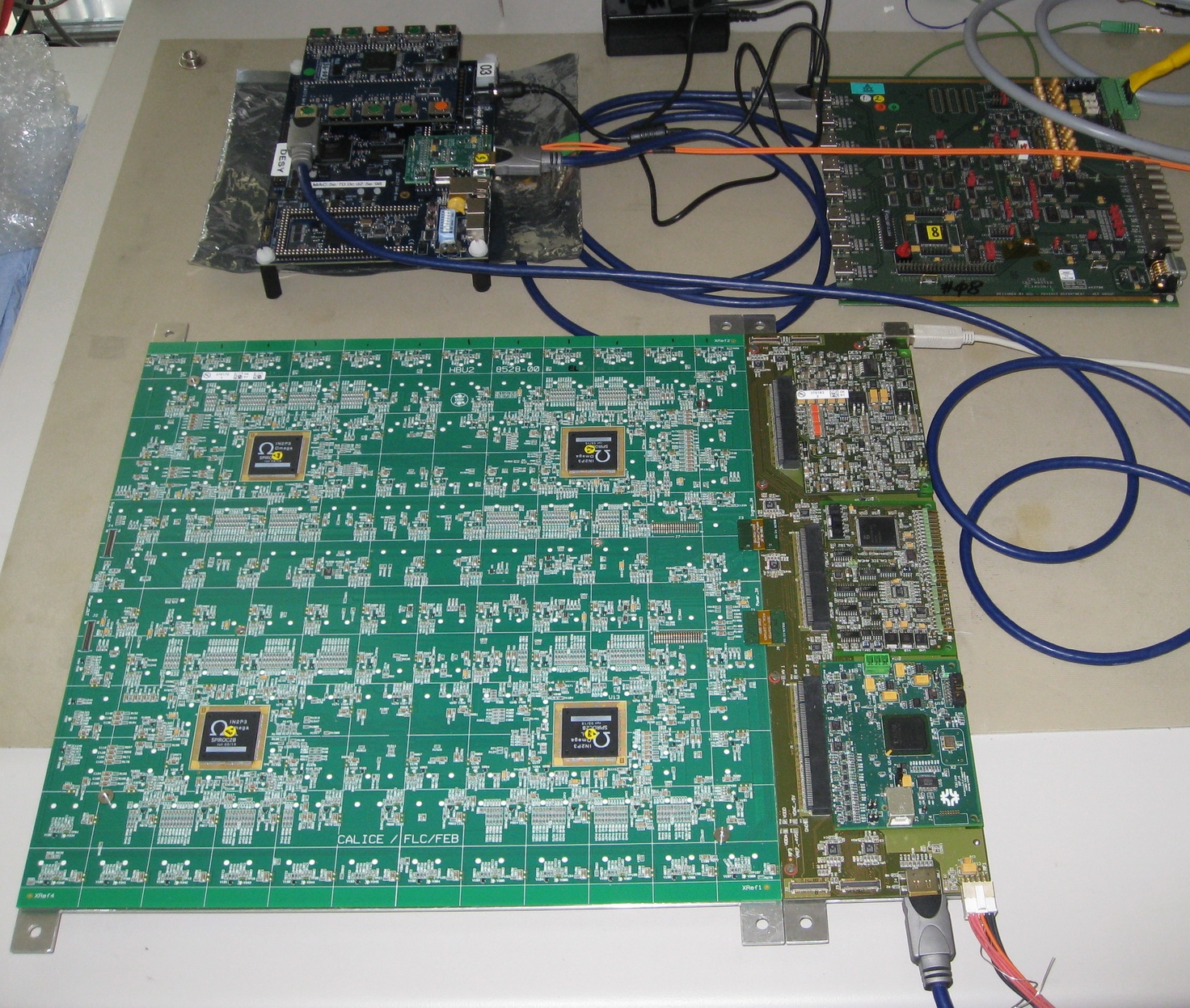}}
\caption{Second version of the HCAL base unit connected to the Central
  Interface Board that hosts the DIF (bottom), CALIB (middle) and
  POWER (top) boards. The connection to the Link/Data Aggregator (top
  left) as well as to the Clock and Control Card (top right) is also
  shown.}
\label{DAQsetup}
\end{figure}

A second version of the HBU has been produced (using the SPIROC2b
chip) and first tests of its functionality have started. It has an
optimized electronics design and especially the implementation of the
calibration system has been updated. Figure~\ref{DAQsetup} shows the
new HBU as it is connected to the DAQ interface modules (see
Sec.~\ref{DAQ}).

\subsection{Calibration system} \label{calib}

\begin{wrapfigure}{r}{0.4\columnwidth}
\centerline{\includegraphics[width=0.35\columnwidth]{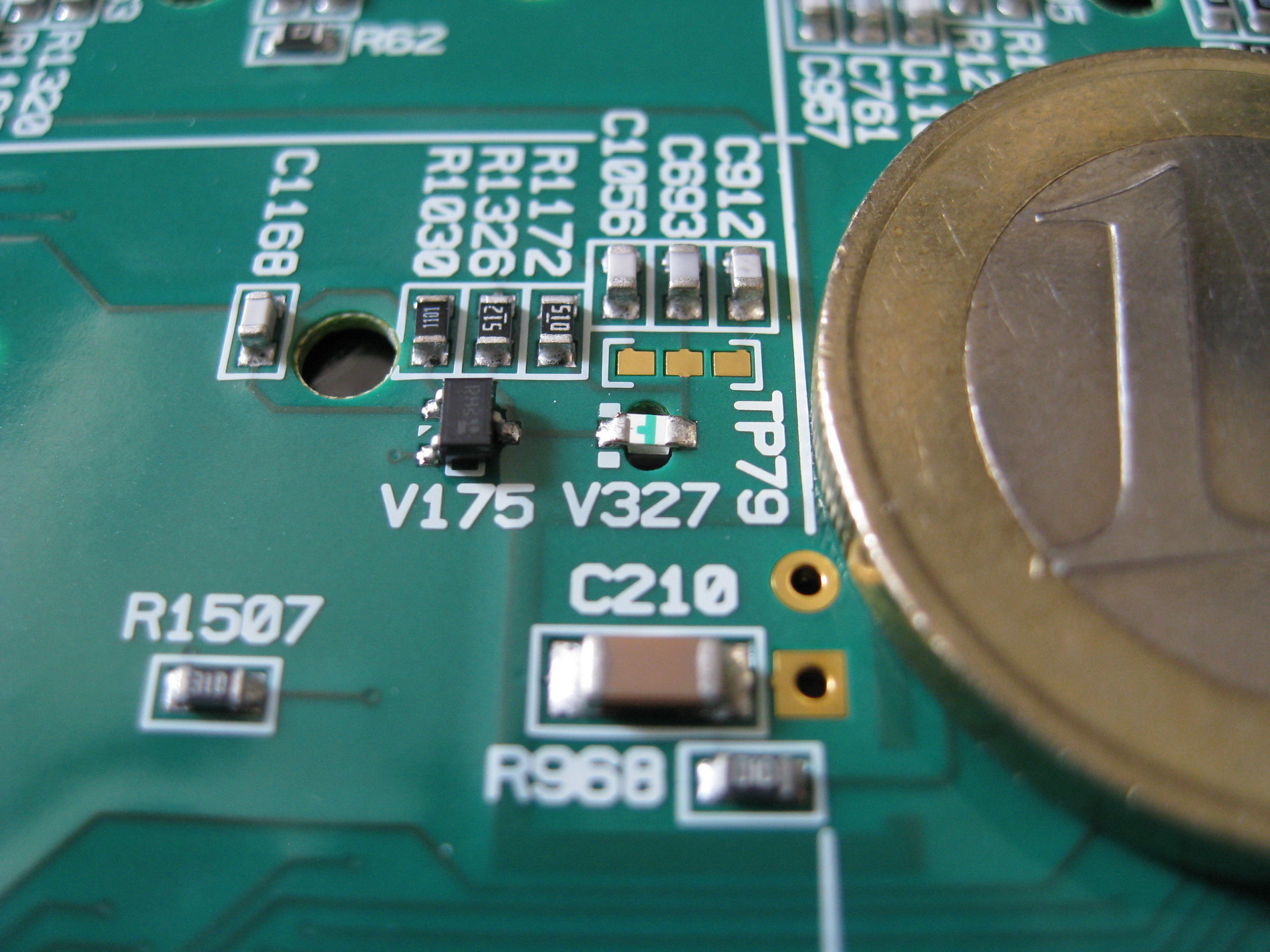}}
\caption{Light calibration system of the new HBU design with one
  integrated LED per channel.}
\label{LED}
\end{wrapfigure}

The response of the SiPM is strongly temperature and bias voltage
dependent and it saturates for high light intensities due to the
limited number of pixels. Therefore, a gain-calibration and
saturation-monitoring system with high dynamic range is needed. For
the gain calibration a low light intensity is used to measure the gain
of the photo detector from the distances of individual peaks in a
single-pixel-spectrum, while for high light intensities, that
correspond to $\sim$100 minimum-ionizing particles, the SiPM
saturates. Two concepts are currently under under study:
\begin{itemize}
\item Each channel features an integrated LED, that is reverse mounted
  into a hole in the PCB and shines on the tile. This system is used
  in the HBUs in the DESY test setups, see Fig.~\ref{LED}.
\item Few strong LEDs that are placed on a special board outside the
  detector, while the light is distributed via notched
  fibers~\cite{Jiri}.
\end{itemize}
Both options have been tested successfully in the DESY test beam
environment. For the design of the new HBUs the driver circuit of the
integrated LED system has been optimized in order to achieve optical
pulse lengths of about 10\,ns for a wide range of amplitudes. This is
needed to provide a good quality for the single-pixel-spectra that can
be evaluated in the gain calibration procedure with high efficiency.
UV LEDs have been used for this purpose, while the option of blue LEDs
is kept for future developments. It has also been shown that the light
intensity of the UV LEDs for a bias voltage of up to 10\,V is high
enough to saturate the SiPMs. In order to minimize the number of
calibration runs in test beam operations, it is important to have a
reasonable uniform light output for a large sample of LEDs. This is
achieved by using an array of different LED loading capacities that
can be chosen individually channel by channel, since all channels have
a common LED bias voltage.

\subsection{Data acquisition interface} \label{DAQ}

As can be seen in Fig.~\ref{halfsector}, each slab of HBUs is
connected to a DAQ system at the end of each layer~\cite{Reinecke}.
The Central Interface Board (CIB) hosts the Detector Interface (DIF),
the steering board for the calibration system (CALIB) and the power
module (POWER), that distributes all voltages needed in the slab. The
DIF serves as the interface between the inner-detector ASICs and the
DAQ. The middle slab in each layer is directly connected to the CIB
via flexleads, while the side slabs are connected to the CIB via side
interface boards (SIBs). There are currently two options for the
operation of the DAQ system. For laboratory tests a USB interface is
used that connects the DIF directly to a PC, where a Labview based
user interface is running. This interface has been used for the
measurements presented later in this report. For the operation of a
larger detector a more advanced DAQ system is needed that uses the
Link/Data Aggregator (LDA, also shown in Figs.~\ref{halfsector}
and~\ref{DAQsetup}) that collects the data from the DIFs to send them
via an optical link to the DAQ PCs and the Clock and Control Card
(CCC) that provides the common clocks for all components. All modules
are available at DESY and are currently under test to commission the
communications between the different systems. Figure~\ref{DAQsetup}
shows the complete DAQ setup as it is currently used at DESY,
including the new HBU, the CIB, the LDA and the CCC with all
interconnections.

\section{Measurements and results}

The main task of the current tests is to commission the detector
concept for a larger scale prototype. The functionality of the ASICs
and tiles in the HBU environment has been proven in laboratory and
test beam measurements for the first version of the
HBU~\cite{Reinecke}. Since then many subcomponents have been further
developed and their tests are an ongoing task. Three examples are
reported here.

\subsection{Tiles}

\begin{figure}%[!t]
\centering
\subfigure[] {\includegraphics[width=2.7in]{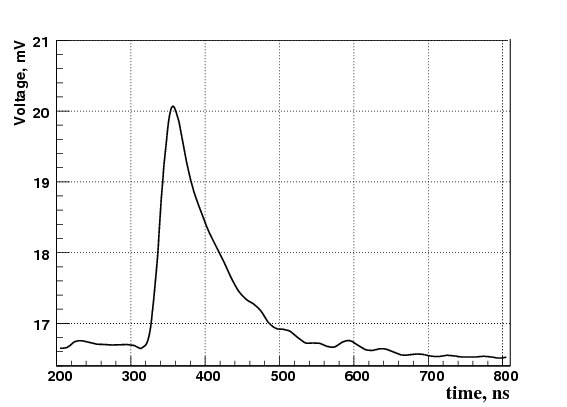}\label{waveform}}
\hspace{1.5cm}
\subfigure[] {\includegraphics[width=2.8in]{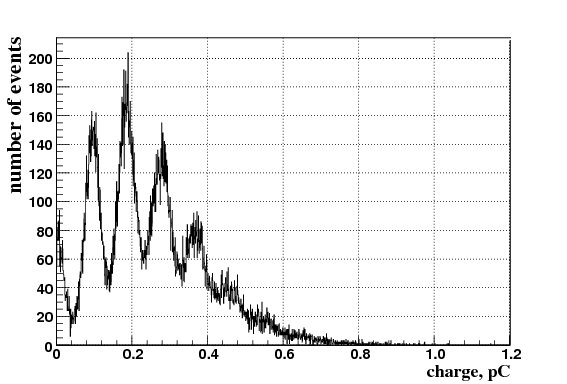}\label{sps}}
\caption{(a) Typical SiPM pulse shape and (b) single-pixel-spectrum
  measured with an integration time of 160\,ns~\cite{Ivan}.}
\end{figure}

For the equipment of the new HBUs a new generation of tiles with SiPMs
is developed. A first sample of 20 tiles has been tested at DESY by
measuring the gains and pulse shapes~\cite{Ivan}. Therefore, the
optimal bias voltage as indicated by ITEP has been applied to each
SiPM and light pulses with differing amplitudes from blue LEDs were
used to illuminate them. A typical pulse shape as measured with a QDC
directly at the output of the SiPM is shown in Fig.~\ref{waveform}.
The mean value of the pulse lengths, defined as the widths of the
signals at 10\% of the maximal amplitude, is $\sim$219\,ns, while the
distribution ranges from $\sim$120\,ns to $\sim$320\,ns~\cite{Ivan}.

As discussed in Sec.~\ref{calib}, the gain of each SiPM can be
calibrated by measuring a single-pixel-spectrum. All tested tiles
could be calibrated with charge integration times between 50\,ns and
160\,ns. Note that the SiPM signal is passed through a shaper with
50\,ns shaping time in the ASIC and that this corresponds to roughly
100\,ns integration time when using a QDC. A typical
single-pixel-spectrum is shown in Fig.~\ref{sps}. The measured gains
vary from $0.4\cdot 10^6$ to $0.8\cdot 10^6$, defined for an
integration time of 160\,ns. Assuming an integration time of 50\,ns
the gain is a factor of $\sim$2.6 smaller~\cite{Ivan}. A larger sample
of tiles will be needed and tested in the near future when the new
HBUs have to be fully equipped.

\subsection{First SPIROC2b measurements in HBU environment}

\begin{figure}[!b]
\centering
\subfigure[] {\includegraphics[width=2.9in]{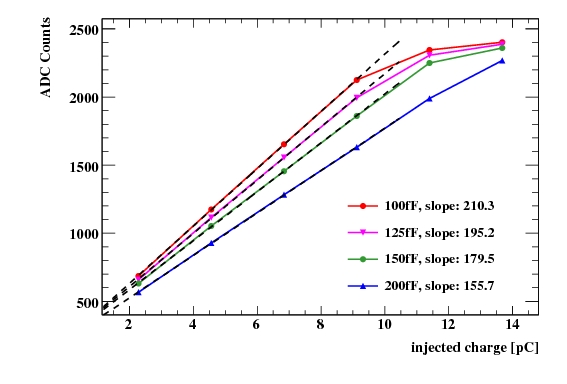}\label{HGlinearity}}
\hspace{1.5cm}
\subfigure[] {\includegraphics[width=2.9in]{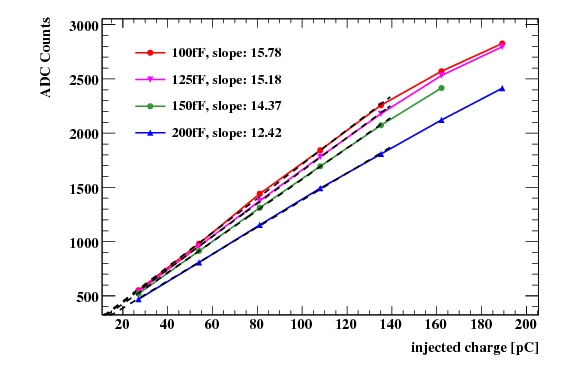}\label{LGlinearity}}
\caption{Linearity test of the channel-wise preamplifier gain
  selection for one example channel and different settings for the
  preamplifier feedback capacitance for the (a) high and (b) low gain
  channels~\cite{Oskar}.}
\label{linearity}
\end{figure}

The SPIROC2b is the latest version of the SPIROC chip family and has
been mounted on the first version of the HBU for first system tests.
The preamplifier gain can be adjusted individually for every readout
channel in order to compensate fluctuations in the photo detector
gain. The preamplifier feedback capacitance can be varied in the range
between 25\,fF and 1575\,fF in steps of 25\,fF, while the ratio of the
high gain and low gain amplifications is fixed to 10. The linearity
and the overall behaviour has been tested with charge injection for
multiple channels and multiple values for the capacitors. It is
expected that a value between 100\,fF and 200\,fF is optimal for
efficient use of the dynamic range of the ADC in the low gain channel.
At the same time it is important to get a reasonable resolution in the
high gain channel to measure single-pixel-spectra for the calibration
of the photo detector gain.

Figure~\ref{HGlinearity} shows the measurements of the amplitude in
ADC tics as a function of the injected charge, performed for one
example channel in high gain mode and different settings for the
preamplifier gain. The chip is well linear up to injected charges of
around 12\,pC. It has to be noted that the ratios between the slopes
of the linear fit functions and between the corresponding feedback
capacitances are not equal. Figure~\ref{LGlinearity} shows the same
measurement for the low gain channel. Again, good linearity can be
observed for the whole dynamic range of the ADC.

\subsection{Auto-gain} \label{AG}

\begin{wrapfigure}{r}{0.45\columnwidth}
\centerline{\includegraphics[width=0.4\columnwidth]{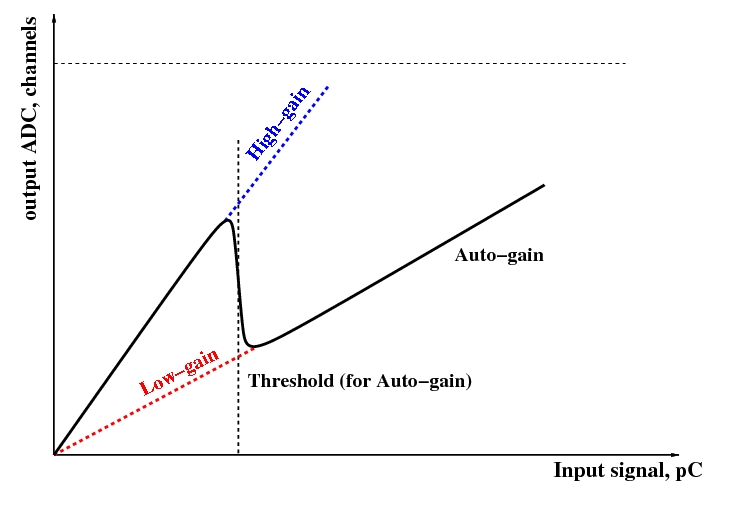}}
\caption{Functionality of the auto-gain mode~\cite{Ivan}.}
\label{Autogain_schema}
\end{wrapfigure}

\begin{figure}[!b]
\centering
\subfigure[] {\includegraphics[width=2.6in]{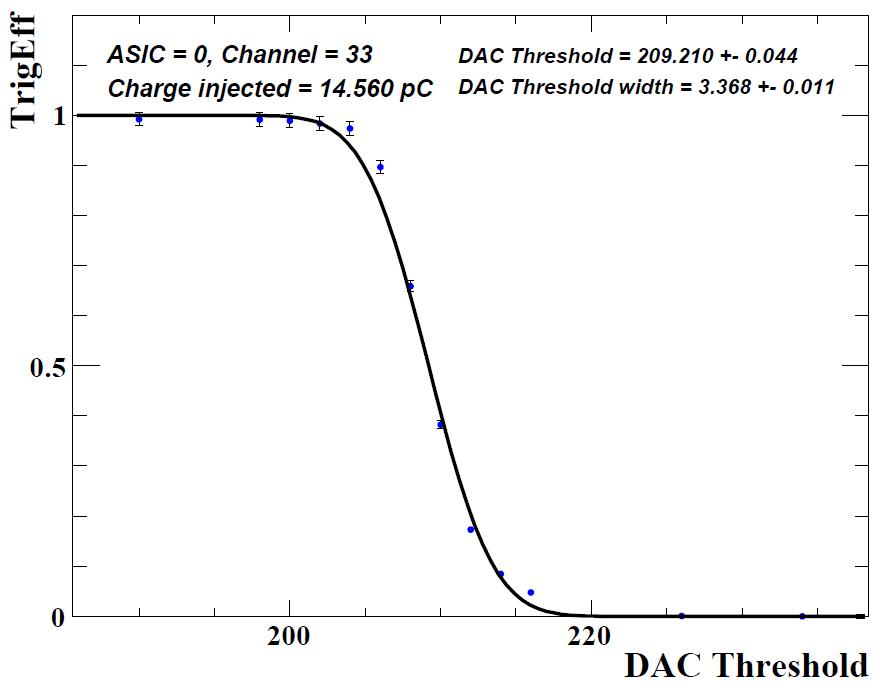}\label{Autogain_scurve}}
\hspace{1.5cm}
\subfigure[] {\includegraphics[width=2.8in, height=2.1in]{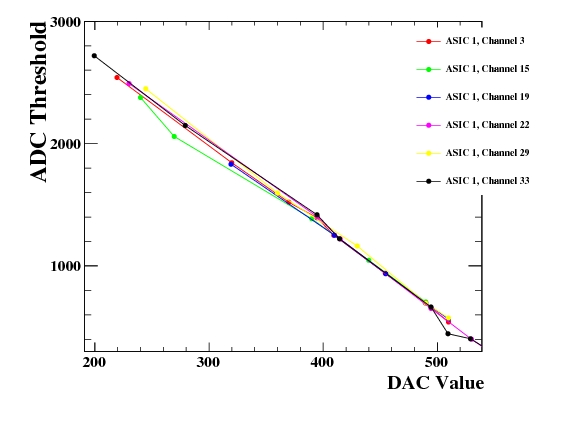}\label{Autogain_ADCvsDAC}}
\caption{(a) Efficiency curve of the auto-gain threshold for a fixed
  injected charge corresponding to the saturation region of the high
  gain mode and (b) position of the threshold as a function of the DAC
  setting~\cite{Ivan}.}
\end{figure}

The SPIROC ASIC can chose the gain channel automatically by comparing
the signal output from a fast shaper to a predefined threshold, which
is set by a 10-bit DAC. If the signal exceeds the threshold, the
amplitude measured in the low gain channel of the chip is stored in
the data, otherwise the amplitude measured in the high gain channel is
stored. An additional bit gives the information about the chosen
channel. Fig.~\ref{Autogain_schema} shows a schematic drawing of the
auto-gain functionality, which has been tested extensively at DESY
with charge injection and LED light for the SPIROC2 ASIC~\cite{Ivan}.
It is expected that the auto-gain efficiency curve, extracted from a
threshold scan at a fixed input charge, has a non-zero width. One
example measurement is shown in Fig.~\ref{Autogain_scurve}. The amount
of injected charge corresponds roughly to the beginning of the high
gain saturation region, i.e. the region where a switching from high
gain mode to low gain mode is desirable. The width of the efficiency
curve in this region corresponds to approximately two photo electrons,
which is sufficiently narrow. The amplitude of the threshold as a
function of the DAC value has also been measured in order to test the
dynamic range of the DAC, which is displayed in
Fig.~\ref{Autogain_ADCvsDAC}. The maximal threshold that can be set by
the DAC corresponds to an injected charge of approximately 20\,pC,
which is well in the saturation region of the high gain mode. Since no
significant shifts or non-linearities of the auto-gain mode are
observed~\cite{Ivan}, it can be used in future test beam measurements.

\section{Summary and outlook}

The CALICE collaboration is currently developing an engineering
prototype of an AHCAL for a future LC experiment. It is foreseen to
build a full detector layer with a realistic readout scheme, including
integrated front-end electronics and a new DAQ system to demonstrate
the feasibility and scalability of the concept. A first version of the
setup has been tested extensively with charge injection, LED light and
electron beams. All results obtained with this setup show very good
results in terms of functionality and performance and lead to an ever
increasing understanding of the system. As an example, the performance
of the auto-gain selection capability has been discussed in this
report.

In order to further develop the prototype, all relevant subcomponents
have been redesigned, including scintillating tiles, ASICs, HBUs, the
LED calibration system and the detector/DAQ interface modules. A
second version of the system has been build and the status has been
presented here. First measurement results for the SPIROC2b ASIC have
been discussed and all tests with charge injection performed so far
show good results.

Two of the most important next steps are the further development of
the new DAQ system and the commissioning of the new HBUs in terms of
overall functionality (including power pulsing) and ASIC performance.
The new subcomponents have to be tested in detail to reestablish a
fully running system. Additionally, the single HBU setup will be
extended to a multi-HBU setup. First, a full calorimeter layer with a
length of 6 HBUs will be tested and finally a multi-layer stack will
be constructed. A possible application is the detailed measurement of
the time structure of hadronic showers.

\section{Acknowledgements}

The author gratefully thanks Erika Garutti, Peter G\"ottlicher, Oskar
Hartbrich, Mathias Reinecke, Felix Sefkow and Ivan Tolstukhin for very
useful discussions and valuable contributions to the results presented
here.

\bibliographystyle{elsarticle-num}
\bibliography{<your-bib-database>}

\end{document}